\documentclass[american,preprint]{revtex4}
\usepackage[T1]{fontenc}
\usepackage[latin1]{inputenc}
\usepackage{babel}
\usepackage{textcomp}
\usepackage{amsmath}
\usepackage{graphicx}
\usepackage[unicode=true,pdfusetitle,
 bookmarks=true,bookmarksnumbered=false,bookmarksopen=false,
 breaklinks=false,pdfborder={0 0 1},backref=false,colorlinks=false]
 {hyperref}
\usepackage{breakurl}

\makeatletter
\@ifundefined{textcolor}{}
{%
 \definecolor{BLACK}{gray}{0}
 \definecolor{WHITE}{gray}{1}
 \definecolor{RED}{rgb}{1,0,0}
 \definecolor{GREEN}{rgb}{0,1,0}
 \definecolor{BLUE}{rgb}{0,0,1}
 \definecolor{CYAN}{cmyk}{1,0,0,0}
 \definecolor{MAGENTA}{cmyk}{0,1,0,0}
 \definecolor{YELLOW}{cmyk}{0,0,1,0}
 }

\usepackage{graphicx}
\usepackage{amssymb}

\usepackage{babel}

\makeatother

\begin{document}

\title{A Formalism for Scattering of Complex Composite Structures. 1 Applications
to Branched Structures of Asymmetric Sub-Units.}

\author{Carsten Svaneborg$^{1,2*}$ and Jan Skov Pedersen$^{2}$}

\affiliation{$^{1}$Center for Fundamental Living Technology, Department of Physics,
Chemistry and Pharmacy, University of Southern Denmark, Campusvej 55, DK-5320
Odense, Denmark}

\affiliation{$^{2}$Department of Chemistry and Interdisciplinary Nanoscience
Center (iNANO), University of Aarhus, Langelandsgade 140, DK-8000
Århus, Denmark.}
\begin{abstract}
We present a formalism for the scattering of an arbitrary linear or
acyclic branched structure build by joining mutually non-interacting
arbitrary functional sub-units. The formalism consists of three equations
expressing the structural scattering in terms of three equations expressing
the sub-unit scattering. The structural scattering expressions allows
a composite structures to be used as sub-units within the formalism
itself. This allows the scattering expressions for complex hierarchical
structures to be derived with great ease. The formalism is furthermore
generic in the sense that the scattering due to structural connectivity
is completely decoupled from internal structure of the sub-units.
This allows sub-units to be replaced by more complex structures. We
illustrate the physical interpretation of the formalism diagrammatically.
By applying a self-consistency requirement we derive the pair distributions
of an ideal flexible polymer sub-unit. We illustrate the formalism
by deriving generic scattering expressions for branched structures
such as stars, pom-poms, bottle-brushes, and dendrimers build out
of asymmetric two-functional sub-units.
\end{abstract}
\maketitle

\section{Introduction}

Scattering techniques, such as light scattering, small-angle neutron
or X-ray scattering (LS, SANS and SAXS, respectively) are ideally
suited for probing the structure of suspensions of macromolecules,
colloidal particles, and self-assembled structures, see e.g. \cite{GuinierFournet,HigginsBenoit,LindnerZemb}.
To extract as much structural information as possible, the data obtained
from a scattering experiment need to be analyzed via extensive modeling,
since scattering techniques do not provide a real space picture or
representation of the structure. A prerequisite for the data modeling
is the availability of a large number of expressions for the form
and structure factors corresponding to various geometric models for
the structures. Fitting such expressions to the measured scattering
data allows the structural parameters to be extracted in an reliable
and accurate manner. Fortunately, scattering expressions have been
derived for a large number of model structures see e.g. \cite{JanAnalysis3}.
Significant efforts are often involved when deducing new scattering
expressions for the analysis of complex structures. Hence, it is of
great importance to have a simple formalism for how to combine existing
model structures to generate new scattering expressions for new and
more complex structures.

In the case of regular polymer structures, quite a few expressions
have been derived for example linear chains\cite{Debye}, block copolymers\cite{Leibler80,LeiblerBenoit81},
stars\cite{Benoit53,BerryOrofino64,Burchard74}, dendrimers\cite{BurchardKajiwaraNegeStockmayer84,Hammouda,BorisRubinsteinMM1996},
and bottle-brush polymers\cite{CasassaBerry66,JanAnalysis2}. These
expressions has been derived assuming that a specific structure has
been build out of linear polymer sub-units. The polymer sub-units
are assumed to be non-interacting and described by Gaussian chain
statistics. With these assumptions, the scattering from a structure
can be deduced from the contour length distribution separating pairs
of scatterers within a structure. The challenge is then how to derive
the contour length distribution for a given structure.

The scattering from more complex heterogeneous structures such as
block-copolymer micelles\cite{PedersenGerstenberg,PedersenSvaneborg2002}
can also be derived. Here one has to take into account that the scattering
length density and structures of the core and corona chains, respectively,
are different. The micellar scattering has contributions from pairs
of scatterers in the core, pairs of scatterers in the corona and core,
and pairs of scatterers on the same and different chains in the corona.
The scattering is derived by calculating all the pair-distances between
the scatterers, taking their connectivity and structure into account,
and neglecting interactions between the core and the corona chains.
The micelle models has been generalized to include a radial rod-like
connector between the corona chains and the core surface to account
for chain stretching close to the core\cite{SvaneborgPedersenJCP2000}
and has also been generalized to describe various core geometries\cite{PedersenGerstenberg2}.

The situation becomes significantly more complex when taking the intra
molecular interactions such as excluded volume or Coulombic interactions
into account. These can be studied analytically by conformational-space
renormalization group theory see e.g. \cite{deGennes,Freed,freedJCP1983,Freed1987,biswasJCP1994}.
An alternative is to perform computer simulations of molecular models.
Computer simulations have, for instance, been applied to study the
effects of excluded volume interactions in flexible polymers\cite{WittkopJCP1996},
semi-flexible polymers\cite{PedersenMacromolecules96,PedersenEurophysLett99},
bottle-brush polymers\cite{elliJCP2004,yethirajJCP2006,hsuJCP2008}
micelles\cite{PedersenGerstenberg,SvaneborgPedersenPhysRevE2001,SvaneborgPedersenMM2002}
and star-burst polymers\cite{Carl1996,timoshenkoJCP2002,BallauffLikosACIE2004,EcheniqueACS2009}.
Common for renormalization group theory and computer simulations are
that they can only be applied to specific structures, and there are
no general way to generalize the scattering form factors to predict
the scattering from related structures.

In the dilute solution case, suspended structures will on average
be far apart and their mutual interactions can be disregarded. Then
the scattering is given by the form factor of the single structures.
At higher concentrations the mutual interaction between structures
gives rise to spatial correlations, that can be observed as the emergence
of a structure factor peak effects in the scattering spectrum. Various
approaches such as the Random-Phase Approximation\cite{Benoit} or
sophisticated liquid-state theories such as the PRISM formalism\cite{SchweizerCurroPRL1987,SchweizerCurro}
can predict concentration effects on the scattering. However, both
of these approaches require the form factor as an input.

We present a formalism for predicting the scattering from general
linear and branched structures composed of mixtures of heterogeneous
sub-units with arbitrary functionality. The formalism is exact for
sub-units that are mutually non-interacting, for links that are completely
flexible and applies to structures that do not contain loops. No assumptions
are made regarding the internal structure or interactions within the
sub-units. The central idea of regarding a structure as composed by
non-interacting sub-units or blocks describing have been utilized
previously by D. J. Read and H. Benoit et al.\cite{BenoitHadziioannouMacromolecules88,ReadMacromolecules98,TeixeiraJCP2007}.

Here we derive the formalism for sub-units with arbitrary functionality,
and derive the terms required to use whole structures as sub-units
within the formalism itself. We illustrate the formalism with a diagrammatic
interpretation, that establishes a direct connection between a general
branched structure and the scattering expressions characterizing that
structure. In particular, we derive the scattering that results when
two known structures are joined by a common point. We also illustrate
the formalism by deriving the scattering expressions for an $ABC$
structures build out of arbitrary sub-units, and for an $AB$ structure,
chains, alternating chains, stars, chains of stars, pompoms and dendrimers
build out of asymmetric two-functional units. 

In the present paper, we derive the general formalism, and illustrate
it using complex structures composed of a single sub-unit type, while
in an accompanying paper\cite{cs_jpc_submitted2}, we will review
expressions for a variety of sub-units and derive scattering expressions
for simple structures focusing on that the different ways sub-units
can be joined together. Taken together the two papers allow the scattering
from a large variety of heterogeneous branched structures to be derived
with great ease. When modeling an experimental small-angle scattering
spectrum, one typically starts with a geometric model from which a
form factor can be derived. This process can be quite laborious and
has to be repeated until the model describes the experimental data.
Our vision is to build model structures by joining together well defined
sub-units together until we obtain a good fit to the experimental
scattering data. The present formalism is a first step in this direction
as it 'automates' the process of deducing the form factor of a structure.
Furthermore, within the present formalism it is trivial to change
the connectivity of the structure, add new sub-units, or replace existing
sub-units by sub-units with a different structure.

The paper is structured as follows: In Sect. \ref{sec:Theory} the
formalism is presented, and the diagrammatic interpretation of the
physics is illustrated with a general $ABC$ structure in sect. \ref{sub:Examples}.
We introduce the special case of asymmetric two-functional sub-units
in sect. \ref{sec:Two-functional-sub-units}. Polymers comprise the
most important sub-units, and in sect. \ref{sub:Polymer-sub-unit}
we derive the scattering expressions of a polymeric sub-unit. To illustrate
the formalism in the case of two-functional sub-units we derive the
scattering expressions for chains (sect. \ref{sub:Chain}), stars
and chains of stars and pom-poms (sect. \ref{sec:Stars}) and dendrimers
(sect. \ref{sub:Dendrimer}). Finally, we conclude the paper in sect.
\ref{sec:Conclusions}.

\section{Theory\label{sec:Theory}}

The present theory pertains to the small-angle scattering from arbitrary
sub-units and how to efficiently calculate the scattering spectra
of complex hierarchical structures, that can be build by joining such
sub-units at common points denoted vertices. Assume that the $I$'th
sub-unit is composed of point-like scatterers, where the $j$'th scatterer
in the sub-unit is located at a position ${\bf r}_{Ij}$ and has excess
scattering length $b_{Ij}$. The scattering length describes the interaction
between a scatterer and the incident radiation, which could be light,
X-rays or neutrons depending on the nature of the scattering experiment.
Let ${\bf R}_{I\alpha}$ denote the position of the $\alpha$'th reference
point associated with the $I$'th sub-unit. A reference point is a
potential point for connecting the sub-unit to other sub-units. A
single sub-unit can have an arbitrary number of such reference points
associated with it. While the scattering sites are real physical entities,
the reference points are just practical handles that we imagine are
fixed somewhere on the scattering sub-unit. If the sub-unit is a polymer,
then a natural choice would for instance be to have the two ends as
reference points. Once two or more sub-units are connected at the
same reference point, we refer to it as a vertex in the resulting
structure, e.g. if sub-units $I$ and $J$ are joined at reference
point $\alpha$ then ${\bf R}_{I\alpha}={\bf R}_{J\alpha}$ denotes
the same location in space and vertex in the structure. Here and in
the following capital letters refers to sub-units, lower case letters
refers to scatterers inside a sub-unit, and Greek letters refers to
vertices and reference points.

Small-angle scattering experiments measure pair-correlation functions.
For a given sub-unit $I$, we can define three types of pair-correlation
functions: Between all pairs of scattering sites, between all scattering
sites and a specified reference point $\alpha$, and between two specified
reference points $\alpha$ and $\omega$, respectively. These are
most conveniently stated in the form of the Fourier transforms

\begin{equation}
F_{I}(q)=\left(\beta_{I}\right)^{-2}\left\langle \sum_{j,k}b_{Ij}b_{Ik}e^{i{\bf q}\cdot({\bf r}_{Ij}-{\bf r}_{Ik})}\right\rangle _{I},\label{eq:FI}
\end{equation}

\begin{equation}
A_{I\alpha}(q)=\left(\beta_{I}\right)^{-1}\left\langle \sum_{j}b_{Ij}e^{i{\bf q}\cdot({\bf r}_{Ij}-{\bf R}_{Ij})}\right\rangle _{I},\label{eq:AI}
\end{equation}
and

\begin{equation}
\Psi_{I\alpha\omega}(q)=\left\langle e^{i{\bf q}\cdot({\bf R}_{I\alpha}-{\bf R}_{I\omega})}\right\rangle _{I}.\label{eq:PI}
\end{equation}

The parameter $q$ is the modulus of the scattering vector and it
is defined by the angle between the incident and scattered beam and
the wave length of the radiation. In the following, we will denote
$F_{I}$ the form factor, $A_{I\alpha}$ the form factor amplitude
relative to the reference point $\alpha,$ $\Psi_{I\alpha\omega}$
the phase factor between reference points $\alpha$ and $\omega$,
and $\beta_{I}=\sum_{j}b_{Ij}$ the excess scattering length of the
$I$'th sub-unit. The $\langle\text{\ensuremath{\cdots}}\rangle_{I}$
averages are over internal conformations and orientations of the $I$´th
sub-unit. Due to the orientational average, the Fourier transformed
pair-correlation functions only depend on the magnitude of the momentum
transfer $q$. Here and in the rest of the paper, the form factor,
form factor amplitudes, and phase factors are normalized to unity
in the limit $q\rightarrow0$. 

The physical significance of the three scattering terms are as follows.
The form factor determines the scattering intensity one obtains from
a dilute solution of a single type of sub-units. Here and in the following
we neglect the spatial correlations that occur at high concentrations.
The scattering intensity at a certain momentum transfer $q$ is determined
from the all the interfering waves scattered from the scattering sites
in the sub-unit. In a given conformation of the $I$'th sub-unit,
two sites $j$ and $k$ contribute interfering waves with a phase
shift $i{\bf q}\cdot({\bf r}_{Ij}-{\bf r}_{Ik})$ and an amplitude
given by scattering strength $b_{Ij}b_{Ik}$ which is determined by
the excess scattering lengths of the two sites compared to the scattering
length density of the solution. The resulting intensity is averaged
over all conformations and orientations of the sub-unit to produce
$F_{I}(q)$.

The form factor amplitude is the total amplitude of the waves scattered
from the scattering sites in the sub-unit with the phase shift $i{\bf q}\cdot({\bf r}_{Ij}-{\bf R}_{I\alpha})$
measured relative to a specific reference point and a scattering strength
$b_{Ij}$. The resulting amplitude is averaged over all conformations
and orientations to produce $A_{I\alpha}(q)$. The phase factor measures
the average phase difference between two reference points averaged
over sub-unit conformations and orientations. In the special case
where the distance is fixed, then the phase factor is given by $\Psi_{I\alpha\omega}(q)=\sin(q|{\bf R}_{\alpha}-{\bf R}_{\omega}|)/(q|{\bf R}_{\alpha}-{\bf R}_{\omega}|)$.
Neither the form factor amplitude nor the phase factor can be measured
directly for single sub-units in dilution, but their contributions
to complex structures can be inferred using contrast variation variation
techniques.

\begin{figure}
\includegraphics[width=0.5\columnwidth]{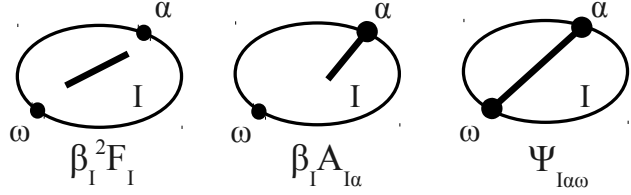}\caption{\label{fig:subunitdiagrams}Diagrams and symbols for the form factor,
form factor amplitude relative to the reference point $\alpha$ and
phase factor between reference points $\alpha$ and $\omega$. The
$\beta_{I}$ prefactor is the total excess scattering length of the
$I$'th sub-unit.}
\end{figure}
The form factor, form factor amplitude and phase factor can be regarded
as propagators of correlation analogous to Feynman diagrams from quantum
field theory or Mayer cluster diagrams from liquid state theory, see
e.g. \cite{QFTBook,LQSBook}. In fig. \ref{fig:subunitdiagrams},
we diagrammatically represent the sub-unit as an ellipse, where the
scattering sites are associated with the inside the ellipse, and reference
points are associated with the circumference of the ellipse. The form
factor is shown as a line inside the sub-unit, because it represents
the sub-units site-to-site pair-correlation function which propagates
position information between unspecified pairs of scattering sites
inside the sub-unit. The form factor amplitude is shown as a line
between a reference point and the sub-unit interior, because it represents
the site-to-reference point pair-correlation function which propagates
position information between scattering sites in the sub-unit and
the specified reference point. Finally, the phase factor is shown
as a straight line between to reference points, because it represents
the reference point-to-reference point pair-correlation function which
propagates position information between two reference points.

\begin{figure}
\includegraphics[width=0.5\textwidth]{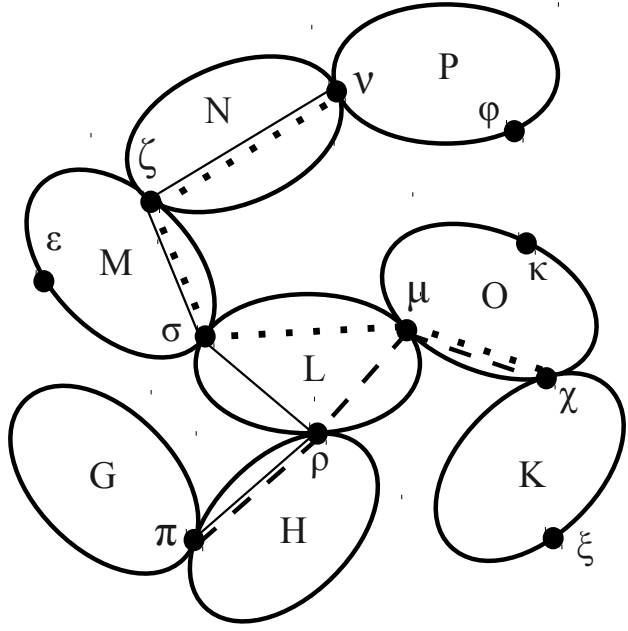}\caption{\label{fig:paths}Diagrammatic representation the connectivity and
vertices ($\nu$, $\rho$, $\zeta$, $\sigma$, $\mu$, $\chi$, $\pi$)
and reference points ($\epsilon$, $\phi$, $\kappa$, $\xi$) of
a star like structure composed of many sub-units. Shown on the figure
are the path $P(\pi,\nu)$ connecting sub-unit $G$ and $P$ (solid),
the path $P(\nu,\chi)$ connecting $P$ and $K$ (dotted), and the
path $P(\pi,\chi)$ connecting $G$ and $K$ (dashed).}
\end{figure}
We can generate complex structures by joining sub-units at reference
points to form vertices in the structure. Since a single sub-unit
can have an arbitrary number of reference points associated with it,
and we can join an arbitrary number of sub-units at a vertex, rather
complex structures can be generated. A diagrammatic example of such
a structure is shown in fig. \ref{fig:paths}, which shows a number
of different sub-units (capital letters) connected by vertices (Greek
letters). The structure has both one, two and three functional sub-units.
Paths that connect specified vertices and reference points through
such structures play a key role in the formalism we present below.
On the figure is shown three examples of such paths connecting vertices.
The structure also contains internal vertices joining sub-units and
external {}``free'' reference points. Further sub-units can be linked
to the {}``free'' reference points and to the internal vertices
of the structure. Note that the same Greek vertex label is used in
all the sub-units reference points linked to a vertex, such that we
get branched structures where each vertex has a unique label. For
a concrete example, polymer sub-units can be joined into a large variety
of block-copolymer and branched structures by end-linking them. Alternatively,
the sub-unit could be a block copolymer, a $f$-functional polymer
star or a $g$'th generation dendrimer, and the more complex sub-units
could be linked tip-to-tip to form chains.

The scattering from such a composite structure $S$ is defined analogously
to the form factor of a single sub-unit as

\begin{equation}
F_{S}(q)\equiv\left(\sum_{I}\beta_{I}\right)^{-2}\left\langle \sum_{I,J}\sum_{i,j}b_{Ii}b_{Jj}e^{i{\bf q}\cdot({\bf r}_{Ii}-{\bf r}_{Jj})}\right\rangle _{S},\label{eq:FT_step1}
\end{equation}
where the first sum is over sub-units. If we assume that joints are
completely flexible such that sub-units joined by a common vertex
can rotate freely with respect to each other, that the resulting branched
structure does not contain any loops, and that all sub-units are mutually
non-interacting, then we arrive at a scattering expression for the
whole structure expressed exclusively in terms of the sub-unit scattering
contributions (eqs. \ref{eq:FI}-\ref{eq:PI}) as

\begin{equation}
=\beta_{S}^{-2}\left[\sum_{I}\beta_{I}^{2}F_{I}(q)+\sum_{\substack{I\neq J\\
\alpha\in I\:\mbox{near}\:\omega\in J
}
}\beta_{I}\beta_{J}A_{I\alpha}(q)A_{J\omega}(q)\prod_{\substack{(K,\tau,\eta)\\
\in\mbox{P}(\alpha,\omega)
}
}\Psi_{K\tau\eta}(q)\right].\label{eq:FT_step2}
\end{equation}

Details of the derivation of this expression is given in the appendix.
The total scattering length of the structure is $\beta_{S}=\sum_{I}\beta_{I}$.
The structural form factor represents the site-to-site pair-correlation
function of the structure build out of sub-units. It consists of two
terms where the first is a sum over all contributions from all pairs
of scatterers inside the same sub-unit, and the second term is double
sum over all the interference contributions from pairs of scatterers
residing in different sub-units. How should the sum in the second
term be evaluated? For each distinct pair of sub-units $I$ and $J$
in the double sum, we identify the vertex $\alpha$ on sub-unit $I$
nearest to $J$ and vertex $\omega$ on sub-unit $J$ nearest to $I$.
Here {}``near'' means in terms of the shortest path originating
at a vertex on $I$ and terminating at a vertex on $J$. We denote
the path connecting $\alpha$ and $\omega$ through the structure
$P(\alpha,\omega)$. For the product, we have to identify all sub-units
$K$ on the path and also identify the vertices $\tau$ and $\eta$
across which the path traverses the sub-unit. This construction is
always unique and well defined for structures that does not contain
loops. While the expression for the structural form factor appears
quite complex, this is mostly due to the notation we have had to introduce
to describe general branched structures. In mathematical terms, structures
such as the one shown in fig. \ref{fig:paths} belong to the class
of hypergraphs since not only can multiple sub-units share the same
vertex, but a single sub-unit can also have multiple reference points.

The form factor expression (eq. \ref{eq:FT_step2}) has a quite simple
physical interpretation. The structural form factor is the pair-correlation
function between all sites in the structure. This is obtained by propagating
position information between all scattering sites in the structure.
When both scattering sites belong to the same sub-unit this is given
by the sub-unit form factors and is described by the first term. The
distance information between scattering sites are on different sub-units
is obtained by propagating position information along paths through
the structure (using eq. \ref{eq:partitioning} in the Appendix).
To propagate site-to-site position information between scatterers
in sub-unit $I$ and scatterers in sub-unit $J$, we first have to
propagate the position information between the scattering sites in
sub-unit $I$ to the vertex $\alpha$ nearest $J$. This is done by
the form factor amplitude $\beta_{I}A_{I\alpha}$. The position information
is then propagated along the path of intervening sub-units towards
the vertex $\omega$ on sub-unit $J$, which is nearest $I$. Each
time a sub-unit is traversed it contributes a phase factor $\Psi_{K\tau\eta}$
to account for the conformationally averaged distance between the
two vertices. Finally the position information is propagated between
the vertex $\omega$ and the scattering sites inside the $J$ sub-unit.
This is done by the final form factor amplitude $\beta_{J}A_{J\omega}$.
Only the amplitudes has a scattering length prefactor, since they
represent the amplitudes of scattered waves from all the scatterers
inside the sub-units relative to the $\alpha$ and $\omega$ vertices
while the product of phase factors represent excess phase contributed
by the path between the vertices. The product of all these propagators
describe the scattering length weighted interference contribution
from the $I$'th and $J$'th sub-units. The same process can be described
in real space, where the product of propagators becomes a convolution
of the site-to-vertex, vertex-to-vertex, and vertex-to-site sub-unit
pair-correlation functions that the propagators represent. This convolution
produces the excess scattering length weighted site-to-site pair-correlation
function for sub-units $I$ and $J$. Since the pair-distances between
$J$ and $I$ also contribute all interference terms are counted twice
in the structural form factor.

In fig. \ref{fig:paths}, we show multiple connected sub-units to
illustrate how to calculate some of the interference contributions
in more detail, and how to find the closest vertices and paths through
a branched structure. For neighbors such as the $G$ and $H$ sub-units,
the vertex $\pi$ on $G$ is nearest $H$, just as it is the vertex
on $H$ nearest $G$. Hence $\alpha=\omega=\pi$, and the path of
sub-units between them $P(\alpha,\omega)=P(\pi,\pi)=\{\}$ is the
empty set. The product over the empty set is unity by definition.
The $G$ and $H$ sub-unit pair contributes a scattering term $\beta_{G}\beta_{H}A_{G\pi}A_{H\pi}$.
For next nearest neighbors such as $M$ and $P$, the vertex $\zeta$
on $M$ is nearest $P$, while $\nu$ on $P$ is nearest $M$. The
path between the two vertices traverses the $N$ sub-unit across the
$\zeta$ and $\nu$ vertices: $P(\zeta,\nu)=\{(N,\zeta,\nu)\}$. The
sub-unit pair $M$ and $P$ contributes an interference term $\beta_{M}\beta_{P}A_{M\zeta}\Psi_{N\zeta\nu}A_{P\nu}$.
For second nearest neighbors such as $N$ and $O$, the path runs
between $\zeta$ and $\mu$, and the path is $P(\zeta,\mu)=\{(M,\zeta,\sigma),(L,\sigma,\mu)\}$.
Hence the sub-unit pair $N$ and $O$ contributes a term $\beta_{N}\beta_{O}A_{N\zeta}\Psi_{M\zeta\sigma}\Psi_{L\sigma\mu}A_{O\mu}$.
Three long paths are shown in the figure. The sub-units $G$ and $P$
contribute a term $\beta_{G}\beta_{P}A_{G\pi}\Psi_{H\pi\rho}\Psi_{L\rho\sigma}\Psi_{M\sigma\zeta}\Psi_{N\zeta\nu}A_{P\nu}$,
the sub-units $P$ and $K$ contribute a term $\beta_{K}\beta_{P}A_{K\chi}\Psi_{O\chi\mu}\Psi_{L\mu\sigma}\Psi_{M\sigma\zeta}\Psi_{N\zeta\nu}A_{P\nu}$,
and the sub-units $G$ and $K$ contribute a term $\beta_{G}\beta_{K}A_{G\pi}\Psi_{H\pi\rho}\Psi_{L\rho\mu}\Psi_{O\mu\chi}A_{K\chi}$.
The reference points $\epsilon$, $\varphi$, $\xi$ , and $\kappa$
can be used to add further sub-units to the structure, but the form
factor is independent of these since no path between vertices will
ever start at, terminate at, or traverse an exterior reference point. 

Note that the assumptions of flexible joints, mutually non-interacting
sub-units, and branched structures without loops allows us to exactly
derive the form factor of the whole structure without making any assumptions
about the specific structure inside the sub-units. In this sense,
the scattering from factor is general as it allows us to write down
scattering expressions for a connected structure without \emph{a priori}
knowledge of the sub-units that the structure is build of. If any
of these assumptions are not strictly fulfilled, then the structural
scattering expression above can be regarded as the zeroth order term
in an perturbation expansion of these effects. Whether this is a good
or bad approximation depends on the detailed structures in the sub-units
and their interactions. A special case of this expression was derived
and used for two functional polymer structures in refs. \cite{BenoitHadziioannouMacromolecules88,ReadMacromolecules98,JanAnalysis3}. 

Using eq. \ref{eq:FT_step2}, we can calculate the form factor for
a whole structure in terms of the fundamental sub-units it is build
of. However, when modeling the scattering from complex structures
it is advantageous to be able selectively modify parts of the structure
while retaining the rest, or to add more sub-units to an existing
structure. To model the resulting change in the scattering spectrum,
it is advantageous also to derive the form factor amplitudes and phase
factors of the whole structure. The form factor amplitude of the whole
structure relative to the reference point or vertex $\alpha$ is defined
as

\begin{equation}
A_{S\alpha}(q)\equiv\left(\sum_{I}\beta_{I}\right)^{-1}\left\langle \sum_{I}\sum_{k}b_{Ik}e^{i{\bf q}\cdot({\bf r}_{Ik}-{\bf R}_{T\alpha})}\right\rangle _{S},\label{eq:AI-1}
\end{equation}
again with the same assumptions as for the form factor of the structure,
we can rewrite express the form factor amplitude of the whole structure
in terms of the sub-unit scattering contributions as

\begin{equation}
=\beta_{s}^{-1}\left[\sum_{\substack{I\\
\omega\in I\,\mbox{near}\,\alpha
}
}\beta_{I}A_{I\omega}(q)\prod_{\substack{(K,\tau,\eta)\\
\in\mbox{P}(\alpha,\omega)
}
}\Psi_{K\tau\eta}(q)\right].\label{eq:Amplitude}
\end{equation}

Again, we refer to the Appendix for details. Here the sum denotes,
that on each sub-unit $I$, we have to identify the vertex on the
sub-unit $\omega$, that is nearest $\alpha$ in terms of structural
connectivity. For each vertex a structure has, there will be a corresponding
form factor amplitude. We can also define and derive the phase factor
between any two vertices on the whole structure as

\begin{equation}
\Psi_{S\alpha\omega}(q)\equiv\left\langle e^{i{\bf q}\cdot({\bf R}_{T\alpha}-{\bf R}_{T\omega})}\right\rangle _{S}=\prod_{\substack{(K,\tau,\eta)\\
\in\mbox{P}(\alpha,\omega)
}
}\Psi_{K\tau\eta}(q).\label{eq:Phase}
\end{equation}

The structural form factor amplitude and phase factor also have simple
physical interpretations despite their complex appearance. The structural
form factor amplitude propagates position information between the
vertex or reference point $\alpha$ and all scattering sites in the
structure. For a particular sub-unit $I$ and vertex $\omega$ nearest
$\alpha$, we first first have to propagate position information between
$\alpha$ and the end $\omega$ vertex on the path path $P(\alpha,\omega)$.
Each sub-unit the path traverses contributes a phase factor $\Psi_{K\rho\sigma}$.
Then position information is propagated between the $\omega$ vertex
and all the sites inside the $I$'th sub-unit, which is represented
by the form factor amplitude $A_{I\omega}$. The product of all the
propagators produces the form factor amplitude which represent the
total amplitude of scattered waves from scattering sites in the structure
measuring the phase relative to the $\alpha$ reference point or vertex.
The same process can be described In real space, where the product
of propagators becomes a convolution of the vertex/reference point-to-vertex
and vertex-to-site sub-unit pair-correlation functions that the propagators
represent. This convolution produces the excess scattering length
weighted site-to-vertex/reference point pair-correlation function
for the whole structure. The interpretation of the structural phase
factor is that it propagates distance information between the vertex/reference
point $\alpha$ and the vertex/reference point $\omega$. This is
the product of all the phase factors $\Psi_{K\rho\sigma}$ for each
sub-unit that has to be traversed on the path $P(\alpha,\omega)$
through the structure. The phase factor represent the phase difference
between two reference points or vertices in the structure averaged
over the conformation and orientations of all sub-units in the structure.

Having defined and derived the form factor amplitudes and phase factors
of a general structure, we have completed the formalism in the sense
that we can now regard any structure described by the formalism as
a sub-unit in the formalism itself. This enables us to compose and
combine several structures to build new structures. For instance,
if a new structure $R$ is formed by joining two known structures
$S$ and $T$ at a common vertex $\alpha$, then the form factor,
form factor amplitude of the resulting structure can be derived from
eqs. \ref{eq:FT_step2}, \ref{eq:Amplitude}. They are given by

\begin{equation}
F_{R}=\beta_{R}^{-2}\left(\beta_{S}^{2}F_{S}+\beta_{T}^{2}F_{T}+2\beta_{S}\beta_{T}A_{S\alpha}A_{T\alpha}\right),\label{eq:FRST}
\end{equation}

\begin{equation}
A_{R\omega}=\beta_{R}^{-\text{1}}\left(\beta_{S}A_{S\omega}+\beta_{T}\Psi_{S\omega\alpha}A_{T\alpha}\right)\quad\mbox{if}\quad\omega\in S\label{eq:ARST}
\end{equation}
where a similar expression applies for the form factor amplitude when
$\omega\in T$ with $S$ and $T$ interchanged. Here the excess scattering
length of the whole structure is $\beta_{R}=\beta_{S}+\beta_{T}$.
The phase factors of the resulting structure $R$ are given by

\begin{equation}
\Psi_{R\rho\sigma}=\begin{cases}
\Psi_{T\rho\alpha}\Psi_{S\alpha\sigma} & \quad\mbox{if}\quad\rho\in T,\sigma\in S\\
\Psi_{S\rho\alpha}\Psi_{T\alpha\sigma} & \quad\mbox{if}\quad\rho\in S,\sigma\in T\\
\Psi_{S\rho\sigma} & \quad\mbox{if}\quad\rho\in S,\sigma\in S\\
\Psi_{T\rho\sigma} & \quad\mbox{if}\quad\rho\in T,\sigma\in T
\end{cases}.\label{eq:PRST}
\end{equation}

These expressions also apply in the special case where one or both
of the structures are single sub-unit, hence they allow us to grow
the scattering expressions for a structure by progressively growing
the structure one sub-unit or sub-structure at the time. Another operation
is to delete a sub-unit $K$ from a given structure, the simplest
way is to collapse all vertices to a single non-scattering point:
this is done by the substitution $F_{K}=0$,$A_{K\cdot}=0$,$\beta_{K}=0$
and $\Psi_{K\cdot\cdot}=1$ where $\cdot$ denotes any vertex of the
$K$'th sub-unit.

What have we learned by this exercise? We have expressed the three
scattering expressions for a whole structure (eqs. \ref{eq:FT_step2},
\ref{eq:Amplitude}, and \ref{eq:Phase}) in terms of three scattering
expressions for the sub-units (eqs. \ref{eq:FI}, \ref{eq:AI}, \ref{eq:PI})
that the structure is composed of. The structural scattering expressions
are generic in the sense that they has been formulated without making
any assumptions of the internal structure of the sub-units. The price
we pay for this is that the formalism is only exact for mutually non-interacting
sub-units that are joined by completely flexible joints. The formalism
makes it easy to derive the form factors of hierarchies of progressively
more complex structures. We have, for instance, shown how to generate
a new structure by two sub-units together and to delete sub-units.
Repeating these operations allows us to join multiple structures together
and remove parts of the structures again. We can also replace sub-units
by sub-units with another structure, or even identify sub-unit motifs
and replace them by other motifs when they have the same external
vertices. The price we pay to be able to generate structures from
structures is that none of these structures can contain loops. To
summarize, the present formalism allows us to construct scattering
expressions for a large class of structures, and we have formulated
it in a way that allows us to easily write a computer program or use
computer aided algebra programs to construct scattering expressions
and evaluating them for any given structure.

The diagrammatic interpretation of the formalism establishes a direct
mapping between the sub-units and the structural connectivity on the
one hand and on the other hand the algebraic structure of the structural
scattering expressions. If we generate a new structure by any of the
structural transformations above, the diagrammatic interpretation
allows us to directly write down the corresponding algebraic transformation
of the scattering expressions. Having proved the validity of diagrammatic
interpretation by deriving the structural scattering expressions,
we can in essence forget these complicated equations and remember
only their simple diagrammatic interpretation, as it allows to write
down the general structural scattering expressions and apply them
to any structure described by the formalism.

\section{Generic ABC structure\label{sub:Examples}}

\begin{figure}
\includegraphics[width=0.5\columnwidth]{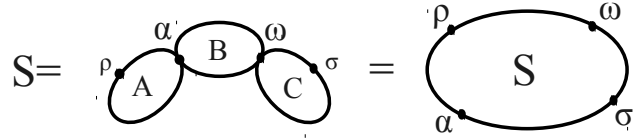}\caption{\label{fig:ABCstructure}A structure $S$ composed of three sub-units
$A$, $B$, and $C$ that are joined at two vertices $\alpha$ and
$\omega$ and has two reference points $\rho$ and $\sigma$. (left)
The whole structure can equally well be regarded as a single four
functional sub-unit with an internal $ABC$ structure. (right)}
\end{figure}
\begin{figure}
\includegraphics[width=0.5\columnwidth]{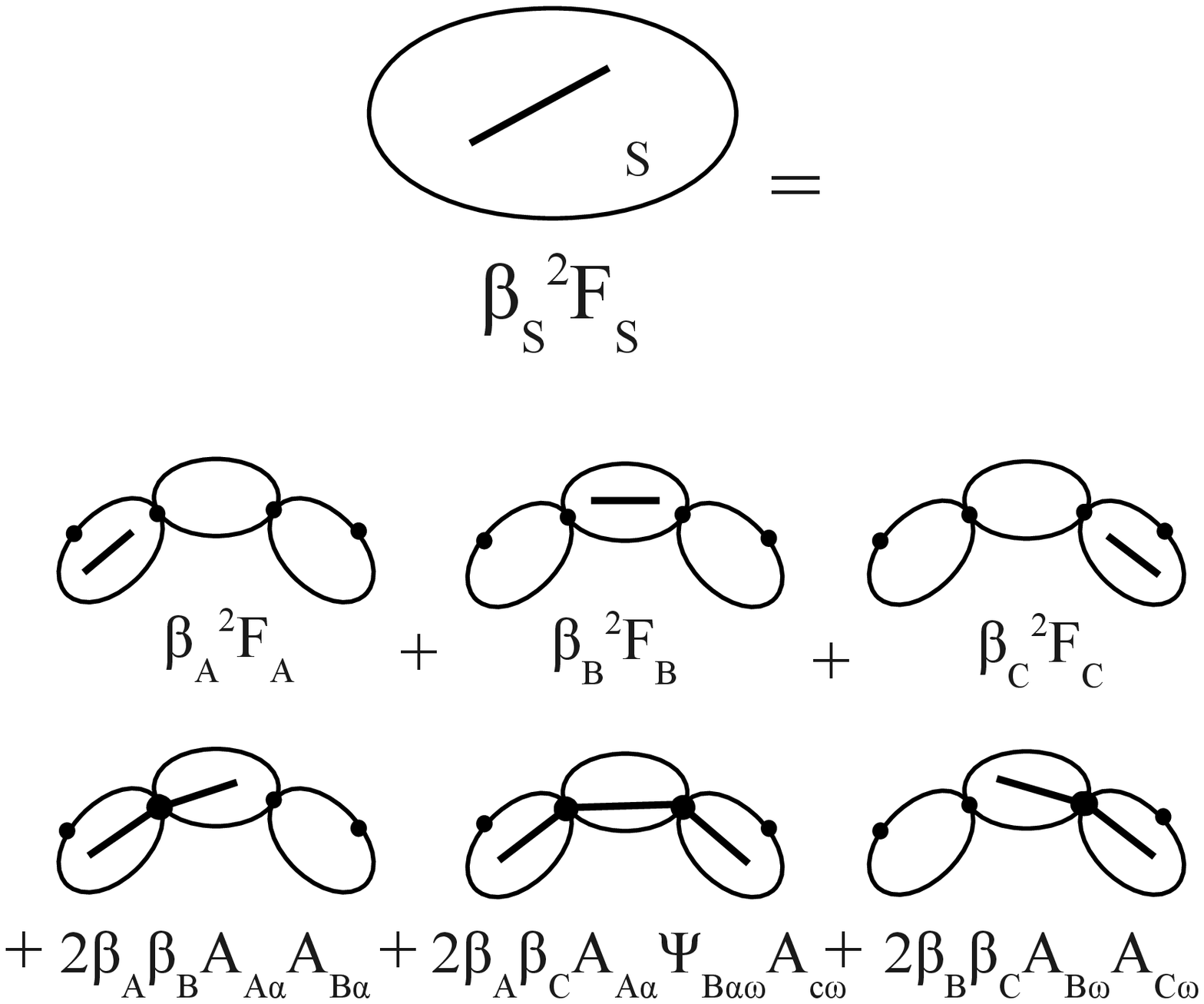}\caption{\label{fig:ABCformfactordiagrams}Form factor of the structure shown
in fig. \ref{fig:ABCstructure} along all the contributing terms and
their diagrammatic representations.}
\end{figure}
\begin{figure}
\includegraphics[width=0.5\columnwidth]{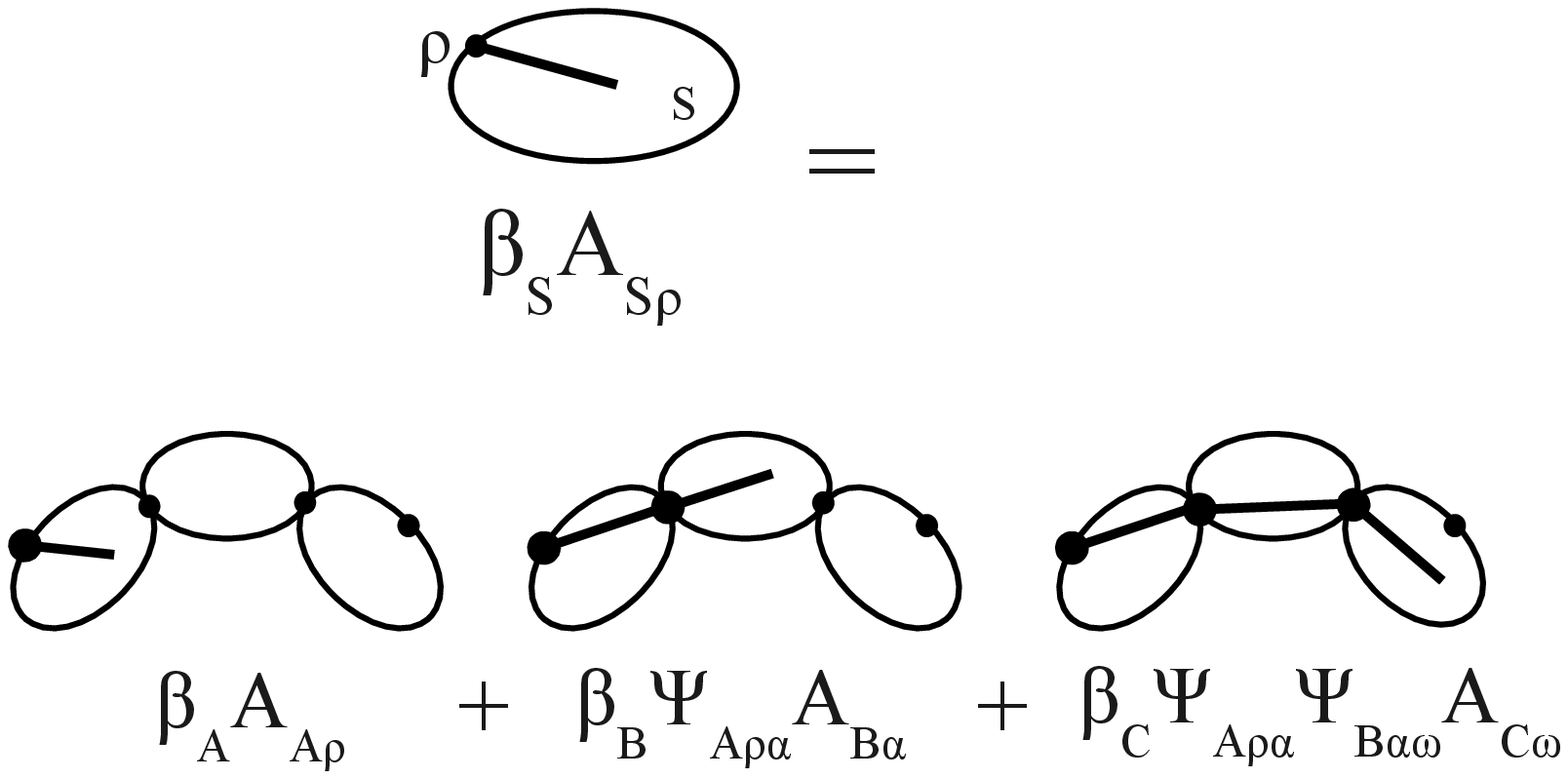}\caption{\label{fig:ABCformfactoramplitudediagrams}Form factor amplitude of
the structure shown in fig. \ref{fig:ABCstructure} relative to the
$\rho$ vertex along with all the contributions and their diagrammatic
representation.
}
\end{figure}

\begin{figure}
\includegraphics[width=0.5\columnwidth]{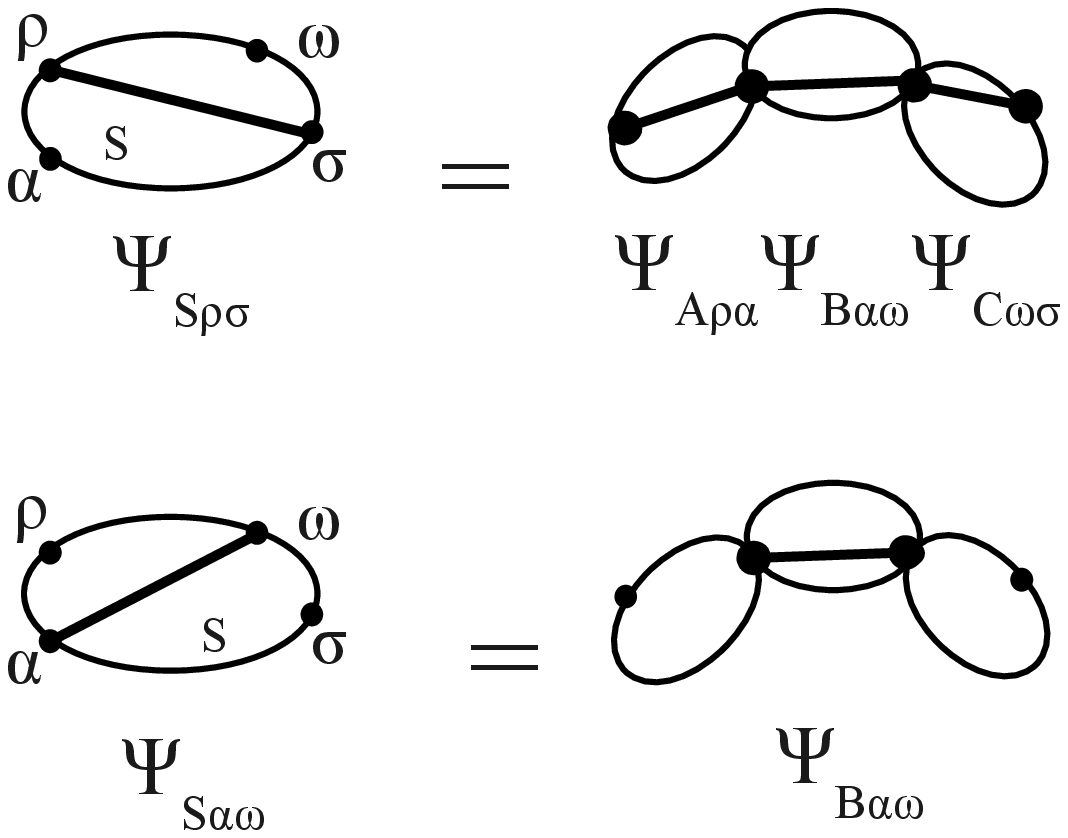}\caption{\label{fig:Phasefactordiagrams}Phase factors the structure shown in fig. \ref{fig:ABCstructure} between the $\rho$ and $\sigma$
vertices and beween the $\alpha$ and $\omega$ vertices.
}
\end{figure}

In fig. \ref{fig:ABCstructure}, we shown the simple $ABC$ structure
where three sub-units have been joined by two vertices to form a linear
chain. This could for instance describe a tri-block copolymer if the
three sub-units are polymers, but it could also describe two colloidal
particles bridged by an adsorbed polymer, or a protein with two dangling
tails. Since we have both eq. \ref{eq:FT_step2}, \ref{eq:Amplitude},
and \ref{eq:Phase} we can derive all the scattering terms describing
the structure, and hence regard the whole structure as a single sub-unit
$S$ characterized by four vertices/reference points. In Fig. \ref{fig:ABCformfactordiagrams}
we show all the terms that contribute to the form factor, in fig.
\ref{fig:ABCformfactoramplitudediagrams} we show all the terms that
contribute to the form factor amplitude relative to $\rho$, and
in fig. \ref{fig:Phasefactordiagrams} we show two of the phase factors
characterizing the $ABC$ structure. Similarly we could
specify the form factor amplitudes relative to all other reference points
or vertices and phase factors between all reference points or vertices.
The diagrammatic representation illustrates how all possible
pair-distances between the sub-units are accounted for by similar
terms in the scattering expressions for the structure. In a similar
way, we can directly write down the generic scattering expressions
for any branched structure of sub-units that we can imagine.

\section{Two functional sub-units\label{sec:Two-functional-sub-units}}

\begin{figure}
\includegraphics[width=0.5\columnwidth]{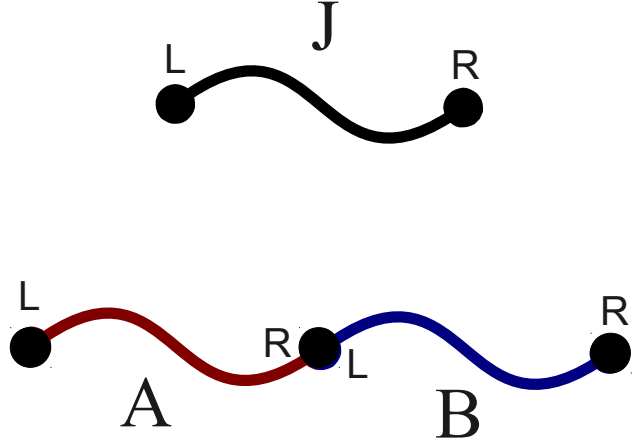}

\caption{\label{cap:linear}A linear sub-unit $J$ shown as a line connecting
the left and right ends (denoted $L$ and $R$). Also shown is an
$AB$ structure where the right end of $A$ is linked to the left
end of $B$.}
\end{figure}
In the following, we will focus on how the general formalism can be
applied to derive the scattering expressions for structures build
of sub-units and structures with only two reference points. We refer
to these as the {}``left'' and {}``right'' ends. This is illustrated
in fig. \ref{cap:linear}. The sub-unit could for instance be polymers
or rigid rods. These are symmetric in the sense that the scattering
from a structure is unaffected if we flip the ends of a polymer or
rod sub-unit (assuming a constant scattering length density along
the sub-unit). The sub-units could also be more complex asymmetric
structures such as $AB$ block copolymers, where the scattering from
a structure will change if we flip a $AB$ block-copolymer turning
it into a $BA$ block-copolymer (e.g. $ABC$ and $BAC$ structures
will produce different scattering spectra). We can also regard any
complex branched structure as being effectively two functional, if
we choose two special vertices or reference points in the structure
and regard these as the {}``left'' and {}``right'' ends of the
structure. For instance for a star, we can pick the tips of two arms
as the reference points.

For two functional sub-units, the $J$'th sub-unit is completely characterized
by $F_{J}$, $A_{JR}$, $A_{JL}$, and $\Psi_{J}$ which denote the
form factor, form factor amplitude relative to the left/right ends
and phase factor between the left and right ends of the sub-unit,
respectively. In the following, instead of using Greek indices for
reference points/vertices we will instead use {}``left'' and {}``right''
ends denoted by subscripts $L$ and $R$ as illustrated in fig. \ref{cap:linear}.
The scattering from an $AB$ structure where the right end of $A$
is joined with the left end of $B$ is a special case of eqs. (\ref{eq:FRST}-\ref{eq:PRST}).
The scattering from an $AB$ structure is given by

\begin{equation}
F_{AB}(q)=\left(\beta_{A}+\beta_{B}\right)^{-2}\left[\beta_{A}^{2}F_{A}+\beta_{B}^{2}F_{B}+2\beta_{A}\beta_{B}A_{AR}A_{BL}\right],\label{eq:F2_AB}
\end{equation}

\begin{equation}
A_{AB,L}(q)=\left(\beta_{A}+\beta_{B}\right)^{-1}\left[\beta_{A}A_{AL}+\beta_{B}\Psi_{A}A_{BL}\right],\label{eq:A2L_AB}
\end{equation}

\begin{equation}
A_{AB,R}(q)=\left(\beta_{A}+\beta_{B}\right)^{-1}\left[\beta_{B}A_{BR}+\beta_{A}\Psi_{B}A_{AR}\right],\label{eq:A2R_AB}
\end{equation}

\begin{equation}
\Psi_{AB}(q)=\Psi_{A}\Psi_{B}.\label{eq:P2_AB}
\end{equation}

Here we choose the free end of the $A$ sub-unit as the {}``left''
end, while the free end of the $B$ sub-unit is the {}``right''
end. We can also easily simplify the scattering expressions from the
$ABC$ structure to produce an effective two functional structure
analogously to the $AB$ structure.

\section{Polymer special case\label{sub:Polymer-sub-unit}}

In the present paper, we focus on scattering expressions of complex
structures. The functions that represent the internal structure of
various sub-units are presented in an accompanying paper.\cite{cs_jpc_submitted2}.
However, here we will (re)derive the functions representing a flexible
polymer. Polymers deserves special attention, since they are the most
important building block of a large variety of synthetic branched
molecular structures. A di-block copolymer $AB$ \cite{Leibler80,LeiblerBenoit81}
consists of two polymer molecules $A$ and $B$ linked end-to-end.
The polymers are symmetric, hence the left and right form factor amplitudes
for the sub-units are identical, and we can discard the $L$ and $R$
subscripts in eqs. (\ref{eq:F2_AB}-\ref{eq:P2_AB}).

Single polymer with $n$ monomers can equally well be regarded as
a di-block copolymer of two identical blocks of $n/2$ monomers each
($\beta_{A}=\beta_{B})$. The form factor, form factor amplitude,
and phase factor are dimensionless numbers. Hence they depend not
only on $q$, which has units of reciprocal length, but also a characteristic
length scale of the polymer. Choosing the radius of gyration $R_{g}^{2}=b^{2}n/6$
where $b$ is the step length and $n$ the number of steps, we can
define a dimensionless parameter $x=q^{2}R_{g}^{2}$. The $n$-monomer
long AB polymer is characterized by the three functions $F_{AB}=F_{polymer}(x)$,
$A_{AB}=A_{polymer}(x)$, and $\Psi_{AB}=\Psi_{polymer}(x)$ while
the two $n/2$-monomer long $A$ and $B$ polymers are characterized
by $F_{A}=F_{B}=F_{polymer}(x/2)$, $A_{A}=A_{B}=A_{polymer}(x/2$),
and $\Psi_{A}=\Psi_{B}=\Psi_{polymer}(x/2)$. Requiring the scattering
from the $AB$ polymer is identical to that of the $A$ and $B$ end-linked
polymers of half the number of monomers in eqs. (\ref{eq:F2_AB}-\ref{eq:P2_AB})
produce the following functional equations

\begin{equation}
F_{polymer}(x)=\frac{1}{2}\left[F_{polymer}\left(\frac{x}{2}\right)+A_{polymer}^{2}\left(\frac{x}{2}\right)\right],\label{eq:f_rw_proof}
\end{equation}

\begin{equation}
A_{polymer}(x)=\frac{1}{2}A_{polymer}\left(\frac{x}{2}\right)\left[1+\Psi_{polymer}\left(\frac{x}{2}\right)\right],\label{eq:a_rw_proof}
\end{equation}
and

\begin{equation}
\Psi_{polymer}(x)=\Psi_{polymer}^{2}\left(\frac{x}{2}\right).\label{eq:p_rw_proof}
\end{equation}

The latter equation has the obvious solution $\Psi_{polymer}(x)=\exp(-\alpha x)$.
Here $\alpha$ is a scaling factor that we set to unity without loss
of generality, since it corresponds to a trivial rescaling of $x$.
Using the ansatz $A_{polymer}(x)=1+\sum_{n=1}^{\infty}\alpha_{n}x^{n}$
and Taylor expanding both sides of eq. (\ref{eq:a_rw_proof}) and
applying the same approach to the form factor we obtain the polymer
solutions

\begin{equation}
\Psi_{polymer}(x)=\exp(-x),\quad A_{polymer}(x)=\frac{1-\exp(-x)}{x},\label{eq:rw}
\end{equation}

\begin{equation}
\mbox{and\quad}F_{polymer}(x)=\frac{2[\exp(-x)-1+x]}{x^{2}}.\label{eq:rw2}
\end{equation}

These results we previously obtained by Hammouda\cite{Hammouda} and
Debye\cite{Debye} by performing the conformational and orientational
averages of eqs. (\ref{eq:FI}-\ref{eq:PI}) explicitly for the Gaussian
pair-distance distributions characterizing a random walk. Here these
solutions emerge as a self-consistency check of the present formalism,
when requiring self-similarity of an object with fractal dimension
two when undergoing a particular scaling transformation. Note that
we could not have obtained the form factor expression unless we had
all three structural scattering expressions.

A polymer sub-unit is completely characterized by the triplet of form
factor, form factor amplitudes and phase factors. The scattering expression
for an flexible ABC block-copolymer is obtained by setting $F_{A}(q)=F_{polymer}(R_{gA}q)$,
$A_{A}(q)=A_{polymer}(R_{gA}q)$, and $\Psi_{A}(q)=\Psi_{polymer}(R_{gA}q)$
and similar for the B and C blocks in the generic ABC structure, where
$R_{gA}$, $R_{gB}$, $R_{gC}$ denotes the radii of gyration of the
three blocks in figs. \ref{fig:ABCformfactordiagrams}-\ref{fig:ABCformfactoramplitudediagrams}.
This is a concrete example of how the present formalism allows us
to write down generic structural scattering expressions and choose
which triplets should be used to characterize the internal structure
of the sub-units. Note that these scattering expressions are specific
to polymers, all other scattering expressions in this paper are generic
in the sense that they remain valid irregardless of what sub-unit
structures are chosen.

\section{Chain structures\label{sub:Chain}}

\begin{figure}
\includegraphics[width=0.5\columnwidth]{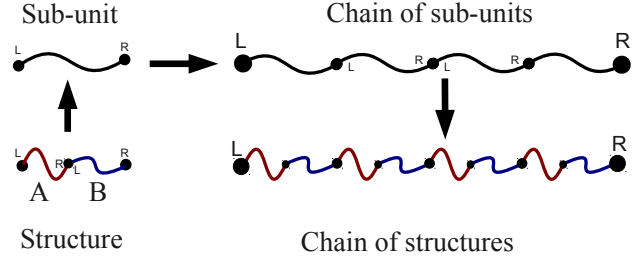}

\caption{\label{cap:abcabcabc}Building a chain of $N$ left-end to right-end
linked sub-units. Substituting the chain sub-unit with any sub-structure
produces a chain of such structures, here is shown the result of substituting
an $AB$ structure to produce a chain of alternating sub-units.}
\end{figure}
Fig. \ref{cap:abcabcabc} shows how a linear chain can be build by
$N$ repeated sub-units joined left end to right end. We regard the
chain as an effective two functional structure defined by the left
and right free ends. In the case of a chain, the scattering expressions
(eqs. \ref{eq:FT_step2}-\ref{eq:Phase}) becomes

\begin{equation}
F_{chain}(q)=\frac{F}{N}+\frac{2A_{R}A_{L}}{N^{2}}\left((N-1)+(N-2)\Psi+(N-3)\Psi^{2}+\cdots+2\Psi^{N-3}+\Psi^{N-2}\right)
\end{equation}

\begin{equation}
=\frac{F}{N}+2\frac{\Psi^{N}-N\Psi+N-1}{N^{2}(\Psi-1)^{2}}A_{R}A_{L}.\label{eq:f_chain}
\end{equation}

The terms of the form $2C_{n}A_{L}\Psi^{n}A_{R}$ encode the fraction
of neighbors pairs $n$ sub-units distant from each other. The form
factor amplitude of a chain relative to the left end is given by

\begin{equation}
A_{chain,L}(q)=\frac{A_{L}}{N}(1+\Psi+\Psi^{2}+\cdots+\Psi^{N-1})=\frac{A_{L}}{N}\frac{\Psi^{N}-1}{\Psi-1}.\label{eq:a_chain}
\end{equation}

Here the prefactors of the $B_{n}A_{L}\Psi^{n}$ encode the fraction
of units that are a certain distance from the vertex $\alpha$. A
identical expression exists for the right end (denoted $R)$ with
$L$ replaced by $R$. Finally the phase factor between the two ends
becomes

\begin{equation}
\Psi_{chain}(q)=\Psi^{N}.\label{eq:p_chain}
\end{equation}

\section{Substitution}

Except for the polymer, all the expressions presented above are generic
in the sense that they have been derived without making any assumptions
about the internal structure of the sub-units themselves. If we, for
instance, insert the polymer scattering expressions into these generic
expressions the result become the specific scattering expressions
for di- and tri-block copolymers and end-linked chains of polymers.
Since we also have derived not only the form factor, but also the
form factor amplitudes and phase factors for these generic structures,
we have completed the formalism, such that we can also use a whole
structure as a sub-unit to build more complex generic structures.
For example, a $ABAB\cdots AB$ structure is a chain of $N$ repeated
$AB$ structures as shown in fig. \ref{cap:abcabcabc}. We can generate
the corresponding generic structural scattering expressions by combining
the $N$-repeated chain expressions (eqs. \ref{eq:f_chain}-\ref{eq:p_chain})
with those of an $AB$ structure (eqs. \ref{eq:F2_AB}-\ref{eq:P2_AB}).
This is done by the substitution $F\rightarrow F_{AB}$, $A_{R}\rightarrow A_{AB,R}$,
$A_{L}\rightarrow A_{AB,L}$, $\Psi\rightarrow\Psi{}_{AB}$ in eq.
\ref{eq:f_chain}. The result for the generic $AB$ $N$-repeated
chain form factor is trivially obtained as

\[
F_{chain,AB}(q)=\frac{1}{N\left(\beta_{A}+\beta_{B}\right)^{2}}\left[\beta_{A}^{2}F_{A}+\beta_{B}^{2}F_{B}+2\beta_{A}\beta_{B}A_{AR}A_{BL}\right]
\]

\begin{equation}
+2\frac{\Psi_{A}^{N}\Psi_{B}^{N}-N\Psi_{A}\Psi_{B}+N-1}{N^{2}\left(\Psi_{A}\Psi_{B}-1\right)\left(\beta_{A}+\beta_{B}\right)^{2}}\left[\beta_{A}A_{AL}+\beta_{B}\Psi_{A}A_{BL}\right]\left[\beta_{B}A_{BR}+\beta_{A}\Psi_{B}A_{AR}\right],\label{eq:F_chain_AB}
\end{equation}
in a similar way, we can easily obtain the left and right form factor
amplitudes and the phase factor for this structure. If we had substituted
$F_{A}\rightarrow F_{chainA}$, $A_{AR}\rightarrow A_{chainA,R}$,
$A_{AL}\rightarrow A_{chainA,L}$, $\Psi_{A}\rightarrow\Psi{}_{chainA}$
and similar for sub-unit $B$ in eq. \ref{eq:F2_AB}, then the result
would have been the generic form factor for an $AA\cdots AABB\cdots BB$
structure. Inserting the polymer expressions (eqs. \ref{eq:rw}-\ref{eq:rw2})
in these generic expressions would specialize them to produce the
corresponding block-copolymer scattering expressions. We could, for
instance, also choose other sub-units e.g. rods and insert the corresponding
scattering expressions for a rod sub-unit to specialize the generic
scattering expression above for a chain of alternating polymers and
rods. Since we can also easily calculate the form factor amplitudes
and phase factor for these structures, these expressions can also
serve as sub-units to build more complex structures out of alternating
chains of sub-units.

\section{Star structures\label{sec:Stars}}

The simplest branched structure is that of a star. The scattering
from a polymeric star was derived by Benoit.\cite{Benoit53,BerryOrofino64,Burchard74}
The generic scattering expression for a star structure of $f$ identical
sub-unit arms attached by their left end to a central point follows
from \ref{eq:FT_step2} as

\begin{equation}
F_{star}(q)=\frac{1}{f}\left[F+(f-1)A_{L}^{2}\right].\label{eq:f_star}
\end{equation}

The star has $fF$ contributions from the form factor of each arm,
while there are $f(f-1)A_{L}^{2}$ interference contributions between
the arms, since the first scatterer can be on any of the $f$ arms,
while the second scatterer can be one of the remaining $f-1$ arms.
We normalize with the total number of contributions which is $f^{2}.$
Each interference term contributes $A_{L}^{2}$ since the two arms
are joined by their left end. Since all arms are joined by the common
center of the star there are no phase factors. 

To make the star an effectively two-functional sub-unit, we can pick
two vertices as the {}``left'' and {}``right'' ends to express
the star form factor amplitude and phase factors. There are two natural
options: The free end of an arm (denoted {}``\emph{e}'') or the
center of the star (denoted {}``\emph{c}''), which leads to three
possible structures center-to-center ({}``cc''), center-to-end ({}``ce''),
or end-to-end ({}``ee''). Their form factor amplitudes and phase
factors are given by

\begin{equation}
A_{star,e}(q)=\frac{1}{f}\left[A_{R}+(f-1)\Psi A_{L}\right],\quad\mbox{{and}\quad}A_{star,c}(q)=A_{L}.\label{eq:a_star}
\end{equation}

\begin{equation}
\Psi_{star,cc}(q)=1,\quad\Psi_{star,ce}(q)=\Psi,\quad\mbox{{and}\quad}\Psi_{star,ee}(q)=\Psi^{2}.\label{eq:p_star}
\end{equation}

\begin{figure}
\includegraphics[width=0.5\columnwidth]{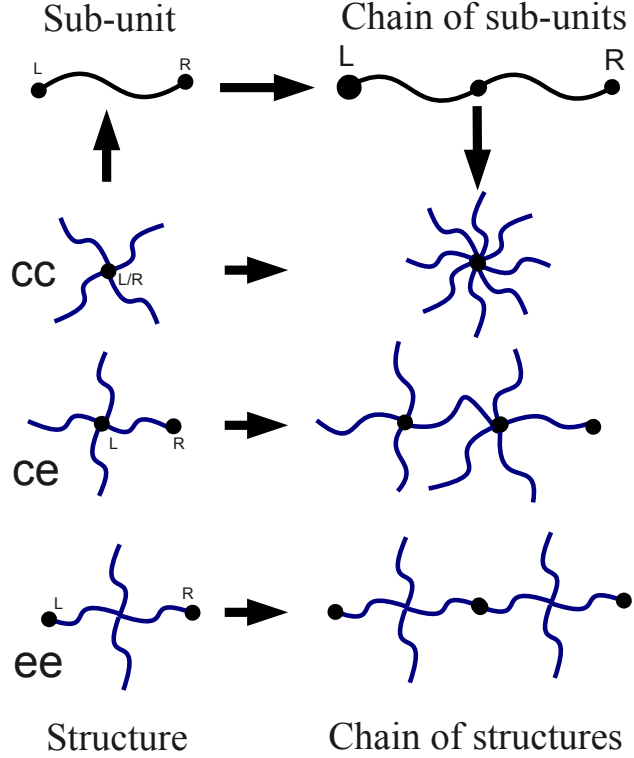}

\caption{\label{cap:bottle-brush}Comparison between the chain structures produced
when substituting the sub-unit for a center-to-center, center-to-end
and end-to-end star.}
\end{figure}

Fig. \ref{cap:bottle-brush} shows the effect of the three substitutions

\[
F\rightarrow F_{star}\quad\quad A_{L}\rightarrow A_{star,c}\quad\quad A_{R}\rightarrow A_{star,c}\quad\quad\Psi\rightarrow\Psi_{star,cc},
\]

\[
F\rightarrow F_{star}\quad\quad A_{L}\rightarrow A_{star,c}\quad\quad A_{R}\rightarrow A_{star,e}\quad\quad\Psi\rightarrow\Psi_{star,ce},
\]
and

\[
F\rightarrow F_{star}\quad\quad A_{L}\rightarrow A_{star,e}\quad\quad A_{R}\rightarrow A_{star,e}\quad\quad\Psi\rightarrow\Psi_{star,ee},
\]
in a chain of $N=2$ repeating sub-units. The result is respectively
a center-to-center, center-to-end and end-to-end linked stars. The
center-to-center linking produces a $Nf$ functional star, while the
center-to-end linking produces a bottle brush with a $f-1$ functional
end and $f+1$ functional branching points inside the bottle brush
and a single arm joining each branch point. End-to-end linking produces
a bottle brush with $f$ functional branch points and where the branch
points are separated by two arms lengths. These structures are geometrically
restricted in the sense that the star centers are separated by zero,
one or two arms lengths. We can obtain more control over the bottle-brush
structure by having a flexible spacer sub-unit between the stars.
Using the $ABAB\cdots AB$ structure with $A$ being a center-to-center
$f$-functional star, and $B$ a spacer sub-unit produces the form
factor of a bottle brush, this is obtained by the following substitutions
in eq. \ref{eq:F_chain_AB}

\[
F_{A}\rightarrow F_{star}\quad\quad A_{AL}\rightarrow A_{star,c}\quad\quad A_{AR}\rightarrow A_{star,c}\quad\quad\Psi_{A}\rightarrow\Psi_{star,cc},
\]

and

\[
F_{B}\rightarrow F_{s}\quad\quad A_{BL}\rightarrow A_{s,L}\quad\quad A_{BR}\rightarrow A_{s,R}\quad\quad\Psi_{B}\rightarrow\Psi_{s}.
\]

Here $F_{s}$, $A_{s,L}$, $A_{s,R}$, $\Psi_{s}$ characterizes the
spacer, which could e.g. be given by the polymer expressions eqs.
\ref{eq:rw}-\ref{eq:rw2}. Since these expressions are obtained by
trivial substitutions, we shall not state them here for sake of brevity.
The scattering from a bottle brush with random positions of the polymeric
side-chains was derived by Casassa and Berry\cite{CasassaBerry66,JanAnalysis2}.
The generic scattering expressions for a miktoarm star structure with
two types of arms can be produced from the $AB$ structure, where
each sub-unit is substituted for stars that are linked center-to-center.
Similarly we can obtain the generic scattering expressions for a
pom-pom structure by specializing the $ABC$ structure (fig. \ref{fig:ABCformfactordiagrams})
to represent two identical center-to-center stars ($A$ and $C$)
separated by a spacer ($B$) sub-unit:

\[
F_{pompom}(q)=\left(2\beta_{star}+\beta_{s}\right)^{-2}\left[2\beta_{star}^{2}F_{star}+\beta_{s}^{2}F_{s}\right.
\]

\begin{equation}
\left.+2\beta_{star}\beta_{s}A_{star,c}(A_{s\text{L}}+A_{sR})+2\beta_{star}^{2}A_{star,c}^{2}\Psi_{s}\right]\label{eq:Fpompom}
\end{equation}
inserting the expressions for the stars eqs. (\ref{eq:f_star}-\ref{eq:p_star}),
and using the polymer expressions eqs. (\ref{eq:rw}-\ref{eq:rw2})
for all sub-units produces the scattering expressions for a pom-pom
polymer. If we want the scattering expression for a pom-pom where
the arms are made of block-copolymers, we use the $ABAB\cdots AB$
structure eq. \ref{eq:F_chain_AB} and the corresponding form factor
amplitudes and phase factor for the arms in the star expressions,
and afterwards insert the polymer expressions for the $A$ and $B$
block copolymers. We could also use a rigid rod for the spacer, or
for all sub-units to produce the scattering expressions for pom-poms
build with rods.\cite{cs_jpc_submitted2} We shall not state the scattering
expressions for any of these specific structures as these any many
others can be obtained by trivial algebraic substitutions of the generic
scattering expressions already given above.

\section{Dendrimer structures\label{sub:Dendrimer}}

\begin{figure}
\includegraphics[width=0.5\columnwidth]{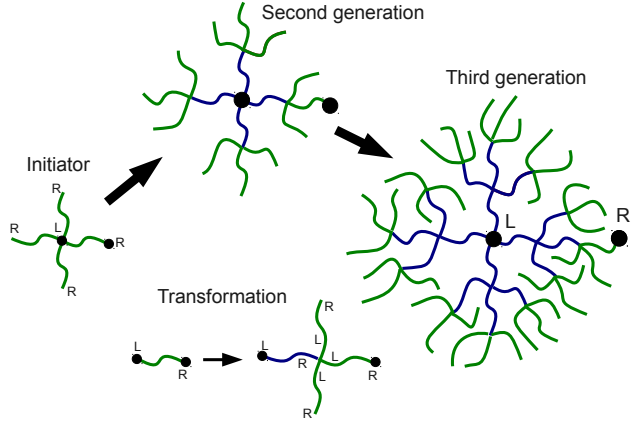}

\caption{\label{cap:Cayley}Generating the $g+1$'th generation Cayley tree
by substituting each leaf (green) in the $g$'th generation Cayley
tree (or an initiator $g=1)$ by a star with one dead branch (blue)
and $f-1$ live leaves (green). Notice how the sub-unit ends are attached
in the Initiator and the Transformation rule. On Initiator, the second,
and third generation dendrimers we also show the two vertices we have
chosen for {}``left'' and {}``right'' ends to make the dendrimer
an effective two functional sub-unit.}
\end{figure}

One dendrimer structure is of particular interest, namely the Cayley
tree. The Cayley tree has a regular structure, where all the branching
points has the same functionality. A Cayley tree structure can be
generated by repeated application of a transformation rule: Starting
from an initiating $f$-functional star, each arm in the initiator
star (or leaf in the dendrimer) is replaced by a $f$ functional star,
where one arm is a {}``dead'' branch connecting to the rest of the
dendrimer and the remaining $f-1$ {}``live'' leaves can grow further.
This is illustrated in fig. \ref{cap:Cayley}.

The scattering functions for a Cayley tree can be generated by performing
by performing the algebraic equivalent of the structural transformation
on the initiator: Starting from the scattering expressions for the
initiator, and replacing the terms characterizing each arm (or leaf)
for the scattering expressions of a $f$ functional star, where each
leaf is converted into a {}``dead'' branch connected to $f-1$ leaf
expressions can be substituted further to generate the next generation.
By starting with an initiator star where all arms are linked to the
center by their left end, and replacing each leaf by a star where
the right end of the dead branch is joined to the left end of the
live leaves we can generate a dendrimer where the left end of all
sub-units connects to the center, and the right end connects to the
periphery of the dendrimer.

The initiator ($I$) is characterized by the following scattering
expressions 

\begin{equation}
F_{I}(q)=f^{-1}\beta_{1}^{2}\left(F_{1}+(f-1)A_{1L}^{2}\right),\quad\quad A_{IL}(q)=\beta_{1}A_{1L},\label{eq:Cayley_initiator1}
\end{equation}

\begin{equation}
A_{IR}(q)=\beta_{1}(A_{1R}+(f-1)\Psi_{1}A_{1L}),\quad\quad\Psi_{I}(q)=\Psi_{1}\label{Cayley_initiator2}
\end{equation}
here $f$ is the functionality of the initiator and $F_{1}$, $A_{1L}$,
$A_{1R}$, $\Psi_{1}$ characterizes the sub-units of the first generation.
The arms in the star are joined by their left ends. For sake of algebraic
simplicity we do not try to keep the expressions normalized. The unnormalized
scattering expressions for the $2$nd generation dendrimer can be
generated by performing the substitutions below with $g=1$ in the
initiator expression, and in general the $g+1$'th generation dendrimer
scattering expression can be generated by applying the substitutions
below in with in the $g$'th generation of the scattering expressions:

\[
F_{g}\rightarrow\beta_{g}^{-2}\left(\beta_{g}^{2}F_{g}+(f-1)\beta_{g+1}^{2}F_{g+1}\right.
\]

\begin{equation}
\left.+2(f-1)\beta_{g}\beta_{g+1}A_{g,R}A_{g+1,L}+2(f-1)(f-2)\beta_{g+1}^{2}A_{g+1,L}^{2}\right)\label{eq:Fdendri}
\end{equation}

\begin{equation}
A_{g,L}\rightarrow\beta_{g}^{-1}\left(\beta_{g}A_{g,L}+(f-1)\beta_{g+1}\Psi_{g}A_{g+1,L}\right)\label{eq:AdendriL}
\end{equation}

\begin{equation}
A_{g,R}\rightarrow\beta_{g}^{-1}\left(\beta_{g}\Psi_{g+1}A_{g,R}+\beta_{g+1}\left[A_{g+1,R}+(f-2)\Psi_{g+1}A_{g+1,L}\right]\right)\label{eq:AdendriR}
\end{equation}

\begin{equation}
\Psi_{g}\rightarrow\Psi_{g+1}\Psi_{g}\label{eq:Pdendri}
\end{equation}

These rules are the algebraic equivalents of the structural transformation
where each {}``live'' $g$-generation leaf is converted into a dead
$g$-generation branch ($g$ terms on the right hand side of the substitution),
and $f-1$ live $g+1$-generation leaves (the $g+1$ terms). The new
leaves are characterized by $F_{g+1}$, $A_{g+1,L}$, $A_{g+1,R}$,
$\Psi_{g+1}$ which allows different sub-units to be used in the various
generations of branches and leafs in the dendrimer. After iterated
substitutions starting from the initiator scattering expressions we
obtain the generic scattering expressions for any generation Cayley
tree. Inserting the polymer expressions (eqs. \ref{eq:rw}-\ref{eq:rw2})
into the expressions will specialize them to a polymeric dendrimer.\cite{BurchardKajiwaraNegeStockmayer84,Hammouda,BorisRubinsteinMM1996}
Alternatively, the $AB$ or $ABAB\cdots AB$ structure (eqs. \ref{eq:F2_AB}-\ref{eq:P2_AB}
or \ref{eq:F_chain_AB}) could be inserted to specialize the dendrimer
scattering expression to provide the scattering expression of a dendrimers
build out of diblock copolymers or copolymers with alternating blocks.

As for the stars, we have also chosen two vertices or reference points
to turn the dendrimer into an effective two functional structure as
shown in fig. \ref{cap:Cayley}. The {}``left'' end is the center
of the dendrimer, while the {}``right'' end is the tip of a leaf.
This allows us to use the dendrimer as a sub-unit, and inserting the
generic dendrimer expressions into the chain will produce the generic
scattering expression for a center-to-end or end-to-end linked chain
of dendrimers.

\begin{figure}
\includegraphics[width=0.5\columnwidth]{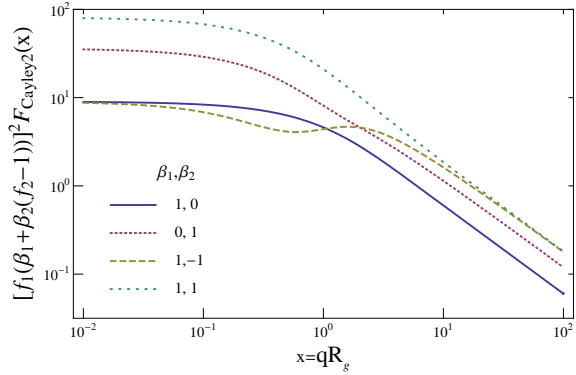}

\caption{\label{cap:CayleyFormfactor-1}Unnormalized form factor of a second
generation Cayley tree with $f=3$ for various choices of contrasts.}
\end{figure}

The effects of varying the contrast for a $3,3$ Cayley tree build
out of identical polymer sub-units is shown in fig. \ref{cap:CayleyFormfactor-1}.
When the solvent matches the outer arms ($\beta_{2}=0$) only the
scattering from the $3$ arms of the central star is shown, while
only the scattering from the $6$ outer arms are shown when the solvent
matches the inner star ($\beta_{1}=0$). At large $x$ this causes
a factor $2$ upwards shift, while at small values of $x$ there are
also interference contributions mediated by the central star. Comparing
$\beta_{1}=\beta_{2}=1$ and $\beta_{1}=1=-\beta_{2}$ we observe
the same scattering at large values of $x$ where the form factors
of the sub-units dominate, but in the latter case the scattering at
$x<1$ is reduced by almost an order of magnitude, since the interference
contributions now reduce the scattering.

\section{Conclusions\label{sec:Conclusions}}

A formalism for calculating the scattering branched structures composed
of arbitrary sub-units of any functionality has been derived and presented.
The formalism allows the scattering from a large class of complex
heterogeneous structures to be derived with great ease. The formalism
is exact in the case where all sub-units are mutually non-interacting,
all sub-unit joins are completely flexible, and the branched structure
does not contain any loops. We have also presented a diagrammatic
illustration of the physical interpretation of the formalism, that
allows us to draw a structure and write down the corresponding scattering
expressions directly. The general formalism was simplified to the
case of two-functional asymmetric sub-units, and illustrated by deriving
generic scattering expressions for AB, ABC, chain structures, alternating
chain structures as well as branched structures such as stars, pom-poms,
bottle-brushes and dendrimers build out of unspecified sub-units. 

A self-consistency requirement was used to derive the scattering expressions
characterizing a polymeric sub-unit, however, none of the structural
scattering expressions derived in the paper makes the assumption that
the structures are build out of polymers. In fact, the scattering
expressions are generic, in the sense that they remain valid irregardless
of the internal structure the sub-units in the structure. In this
sense, the formalism decompose scattering contributions due to the
structural connectivity and due to the sub-unit internal structure.

The scattering contribution due to a sub-unit is completely described
by a triplet of functions: phase factors, form factor amplitudes,
and a form factor. The present formalism provides the triplet of scattering
expressions for a whole structure build out of sub-units. The structural
scattering expressions are complete in the sense that they allow a
composite structure of multiple sub-units to be used as a single sub-unit
within the formalism, which practically means inserting the three
structural scattering expressions recursively into themselves. Complex
hierarchical structures can be build by joining simple sub-units or
complex sub-structures together one by one or by replacing all sub-units
of a certain type by a more complex sub-structure.

A Feynman-like diagrammatic interpretation of the formalism allows
us to map structural transformations to algebraic transformations
of the scattering expressions. In this way, the present formalism
allows us to build complex scattering expressions by simple algebraic
transformations by inserting generic equations representing different
structures into each other, or substituting specific sub-unit triplets
for the three master equations of a sub-structure. We have illustrated
this by deriving the algebraic transformation rules that will produce
the form factor of dendrimer structures.

In an accompanying publication, we will review the sub-unit triplets
of rigid rods, flexible and semi-flexible polymers, polymer loops,
excluded volume polymers. We will also present triplets for thin disks
and spheres, solid spheres and cylinders. We can regard these as infinity-functional
sub-units when joining other sub-units to a random point on their
surfaces. Finally, we will use the formalism presented here to predict
the scattering from structures composed of mixtures of these sub-units.\cite{cs_jpc_submitted2}
We hope that the formalism in the present paper will facilitate the
analysis of experimental scattering data by allowing the scattering
functions to be derived with greater ease for a large variety of complex
structures.

\section{Acknowledgments}

C.S. gratefully acknowledges financial support from the Danish Natural
Sciences Research Council through a Steno Research Assistant Professor
fellowship. C.S. and J.S.P. gratefully acknowledges discussions with
C.L.P. Oliveira.

\part*{Appendix}

Assume that the $I$'th sub-unit is composed of point-like scatterers,
where the $j$'th scatterer in the sub-unit is located at a position
${\bf r}_{Ij}$ and has excess scattering length $b_{Ij}$. Let ${\bf R}_{I\alpha}$
denote the position of the $\alpha$'th reference point associated
with the $I$'th sub-unit. A reference point is a potential point
for connecting the sub-unit to other sub-units. A single sub-unit
can have an arbitrary number of such reference points associated with
it. Once two or more sub-units are connected at the same reference
point, we refer to it as a vertex in the resulting structure. Here
and in the following capital letters refers to sub-units, lower case
letters refers to scatterers inside a sub-unit, and Greek letters
refers to vertices and reference points.

The structural form factor is defined in analogy with the sub-unit
form factor (eq. \ref{eq:FI}) as

\begin{equation}
F_{S}(q)\equiv\left(\sum_{I}\beta_{I}\right)^{-2}\left\langle \sum_{I,J}\sum_{i,j}b_{Ii}b_{Jj}e^{i{\bf q}\cdot({\bf r}_{Ii}-{\bf r}_{Jj})}\right\rangle _{S}.\label{eq:def}
\end{equation}

We can split the double sum into all the diagonal terms with $I=J$
and off diagonal terms with $I\neq J$ and interchange the order of
the conformational averages and sums to produce

\begin{equation}
=\left(\sum_{I}\beta_{I}\right)^{-2}\left(\sum_{I}\left\langle \sum_{i,j}b_{Ii}b_{Ij}e^{i{\bf q}\cdot({\bf r}_{Ii}-{\bf r}_{Ij})}\right\rangle _{S}+\sum_{I\neq J}\left\langle \sum_{i,j}b_{Ii}b_{Jj}e^{i{\bf q}\cdot({\bf r}_{Ii}-{\bf r}_{Jj})}\right\rangle _{S}\right)\label{eq:twoparts}
\end{equation}

If we assume 1) that all sub-units are mutually non-interacting such
that e.g. no excluded volume correlations exists between neighboring
sub-units, 2) that all sub-unit pairs are joined by flexible links,
such that no orientational correlations are be induced by the joints,
and 3) that no loops exists in the structure such that the loop closure
constraints introduces correlations between internal conformations,
then the structural averages can be factorized into products of single
sub-unit averages. Hence we can replace $\sum_{I}\langle\cdots\rangle_{S}$
by $\sum_{I}\langle\cdots\rangle_{I}$ in the first term, which by
eq. \ref{eq:FI} becomes the sum of the sub-unit form factors: $\sum_{I}\beta_{I}^{2}F_{I}(q)$.

\begin{figure}
\includegraphics[width=0.5\textwidth]{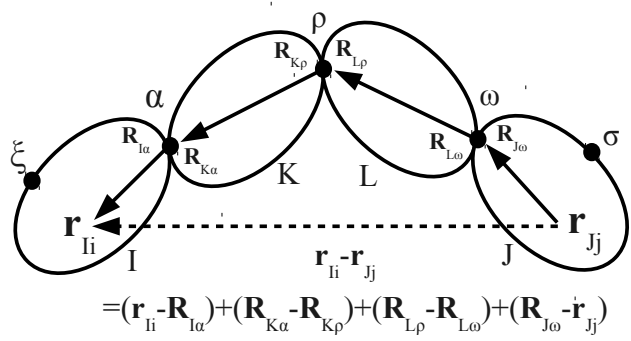}\caption{\label{fig:distances}Vector between a pair of scattering sites ${\bf r}_{Ii}$
and ${\bf r}_{Jj}$ expressed in terms of steps traversing the intervening
sub-units $K$ and $L$ and vertices $\alpha$, $\rho$, $\omega$
between the sub-units $I$ and $J$. Note the reference points of
different sub-units are at identical locations in space because they
are joined: ${\bf R}_{I\alpha}={\bf R}_{K\alpha}$, ${\bf R}_{K\rho}={\bf R}_{L\rho}$,
and ${\bf R}_{L\omega}={\bf R}_{J\omega}$.}
\end{figure}

In the second term, we have weighted pair-distances between sites
in the $I$ and $J$ sub-units. As shown in fig. \ref{fig:distances},
we can always project the vector ${\bf r}_{Ii}$-${\bf r}_{Jj}$ from
the $i$'th scattering site on sub-unit $J$ to the $j$'th scattering
site on sub-unit $I$ onto a path traversing through the structure.
In the structure in fig. \ref{fig:distances}, we first a step from
the $j$'th scattering site to the $\omega$ reference point on sub-unit
$J$, a step through sub-unit $L$ from $\omega$ to $\rho$, a step
through sub-unit $K$ from $\rho$ to $\alpha$, and finally from
the $\alpha$ reference point to the $i$'th scattering site. Since
we have assumed the structure does not contain loops the paths are
always uniquely defined. In general, we have to find the reference
point on sub-unit $I$ which is nearest to $J$ ($\alpha$ in fig.
\ref{fig:distances}), and the reference point on sub-unit $J$ which
nearest sub-unit $I$ ($\omega$ in fig. \ref{fig:distances}). Let
$P(\alpha,\omega)$ denote the path of sub-units and vertices that
has to be traversed between the two ends ($P(\alpha,\omega)=$$\{(K,\alpha,\rho),(L,\rho,\omega)\}$
in fig. \ref{fig:distances}). With these definitions, the general
result for the distance is identity

\begin{equation}
{\bf r}_{Ii}-{\bf r}_{Jj}=\left({\bf r}_{Ii}-{\bf R}_{I\alpha}\right)+\sum_{\substack{(K,\tau,\eta)\\
\in\mbox{P}(\alpha,\omega)
}
}\left({\bf R}_{K\tau}-{\bf R}_{K\eta}\right)+\left({\bf R}_{J\omega}-{\bf r}_{Jj}\right)\quad\mbox{with\quad}\alpha\in I\:\mbox{near}\:\omega\in J.\label{eq:partitioning}
\end{equation}

Inserting this identity into the second term of eq. \ref{eq:twoparts},
and making use of the factorization of the conformational average
into sub-unit averages yields

\begin{equation}
\left\langle \sum_{i,j}b_{Ii}b_{Jj}e^{i{\bf q}\cdot({\bf r}_{Ii}-{\bf r}_{Jj})}\right\rangle _{S}=\left\langle \sum_{i}b_{Ii}e^{i{\bf q}\cdot\left({\bf r}_{Ii}-{\bf R}_{I\alpha}\right)}\right\rangle _{I}\left\langle \sum_{j}b_{Jj}e^{i{\bf q}\cdot\left({\bf R}_{J\omega}-{\bf r}_{Jj}\right)}\right\rangle _{J}\prod_{\substack{(K,\tau,\eta)\\
\in\mbox{P}(\alpha,\omega)
}
}\left\langle e^{i{\bf q}\cdot\left({\bf R}_{K\tau}-{\bf R}_{K\eta}\right)}\right\rangle _{K}\label{eq:monster}
\end{equation}
which by definition eqs. \ref{eq:AI}-\ref{eq:PI} become

\begin{equation}
=\beta_{I}A_{I\alpha}\beta_{J}A_{J\omega}\prod_{\substack{(K,\tau,\eta)\\
\in\mbox{P}(\alpha,\omega)
}
}\Psi_{K\tau\eta}\quad\mbox{with\quad}\alpha\in I\:\mbox{near}\:\omega\in J.\label{eq:interference}
\end{equation}

This is the sub-unit interference scattering contribution which when
inserted in eq. \ref{eq:twoparts} becomes eq. \ref{eq:FT_step2}.
To obtain the form factor amplitude and phase factor of the whole
structure (eqs. \ref{eq:Amplitude} and \ref{eq:Phase}) exactly the
same approach was applied using the following identities for the vector
from a scattering site to a vertex or reference point and between
two vertices or reference points

\begin{equation}
{\bf r}_{Ii}-{\bf R}_{\alpha}=\left({\bf r}_{Ii}-{\bf R}_{I\omega}\right)+\sum_{\substack{(K,\tau,\eta)\\
\in\mbox{P}(\alpha,\omega)
}
}\left({\bf R}_{K\tau}-{\bf R}_{K\eta}\right)\quad\mbox{with\quad}\omega\in J\:\mbox{near}\:\alpha,\label{eq:partitioning-1}
\end{equation}
and

\begin{equation}
{\bf R}_{\alpha}-{\bf R}_{\omega}=\sum_{\substack{(K,\tau,\eta)\\
\in\mbox{P}(\alpha,\omega)
}
}\left({\bf R}_{K\tau}-{\bf R}_{K\eta}\right).\label{eq:partitioning-1-1}
\end{equation}

\section*{Real-space\label{sec:real-space-distances}}

The form factor amplitude is the Fourier transform of the excess scattering
length density distribution around a reference point. Hence we can
directly obtain the (excess scattering length weighted) radial density
distribution of the structure by an inverse Fourier transform as\cite{Pedersen01}

\[
\Delta\rho_{S\alpha}(r)=\int_{0}^{\infty}\mbox{d}q\frac{q\sin qr}{2\pi^{2}r}A_{S\alpha}(q).
\]

We can also relate the structural scattering equations to mean-square
distances of the structure. The radius of gyration has the general
definition: $R_{g}^{2}=-\frac{3}{2}\left.\frac{d^{2}F(q)}{dq^{2}}\right|_{q=0}$.
We can apply a Guinier expansion\cite{Guinier1939} to each sub-unit
as follows

\begin{equation}
F_{I}(q)=1-\frac{\langle R_{Ig}^{2}\rangle q^{2}}{3}+\cdots\quad A_{I\alpha}(q)=1-\frac{\langle R_{I\alpha}^{2}\rangle q^{2}}{6}+\cdots\quad\Psi_{I\sigma\rho}(q)=1-\frac{\langle R_{I\rho\sigma}^{2}\rangle q^{2}}{6}+\cdots.\label{eq:guinier}
\end{equation}

Here $\langle R_{Ig}^{2}\rangle$, $\langle R_{I\alpha}^{2}\rangle$,
and $\langle R_{I\rho\sigma}^{2}\rangle$ denote the radius of gyration,
mean-square distance between all scattering sites and the vertex $\text{\ensuremath{\alpha}}$,
and the mean-square distance between the vertices $\rho$ and $\sigma$
on the $I$'th sub-unit, respectively. The radius of gyration measures
the mean-square distance between unique pairs of sites, hence there
is a factor of two to avoid double counting. The general equation
for the apparent radius of gyration can be obtained by differentiating
eq. (\ref{eq:FT_step2}) in accordance with the Guinier expansions
above, or by factorizing pair-separations into separations along the
sub-units as done for the scattering form factor. The result becomes

\begin{equation}
\langle R_{Sg}^{2}\rangle=\left(\sum_{I}\beta_{I}\right)^{-2}\left\{ \sum_{I}\beta_{I}^{2}\langle R_{Ig}^{2}\rangle+\sum_{\substack{I\neq J\\
\alpha\in I\:\mbox{close}\:\omega\in J
}
}\beta_{I}\beta_{J}\left(\langle R_{I\alpha}^{2}\rangle+\sum_{\substack{(K,\tau,\eta)\in\\
P(\alpha,\omega)
}
}\langle R_{K\tau\eta}^{2}\rangle+\langle R_{J\omega}^{2}\rangle\right)\right\} .\label{eq:rg_struct}
\end{equation}

This expression is analogous to the expression for the form factor,
sub-unit form factors are replaced by their radii of gyration, while
form factor amplitudes are replaced by site-to-end mean-square distances,
and phase factors are replaced by end-to-end mean square distances.
Similar expressions can be deduced for $\langle R_{S\alpha}^{2}\rangle$
and $\langle R_{S\rho\sigma}^{2}\rangle$.

\end{document}